\documentclass{JAC2003}

\usepackage{graphicx}
\usepackage{booktabs}

\begin{document}
\title{COAXIAL WIRE MEASUREMENTS OF FERRITE KICKER MAGNETS}

\author{H. Day\thanks{hugo.day@hep.manchester.ac.uk}, CERN, Switzerland, University of Manchester, UK and Cockcroft Institute, UK\\M.J. Barnes, F. Caspers, E. Metral, B. Salvant, C. Zannini, CERN, Geneva, Switzerland\\
R.M. Jones, University of Manchester, UK and Cockcroft Institute, UK}

\maketitle

\begin{abstract}
Fast kicker magnets are used to inject beam into and eject beam out of the CERN accelerator rings. These kickers are generally transmission line type magnets with a rectangular shaped aperture through which the beam passes. Unless special precautions are taken the impedance of the yoke can provoke significant beam induced heating, especially for high intensities. In addition the impedance may contribute to beam instabilities. The results of longitudinal and transverse impedance measurements, for various kicker magnets, are presented and compared with analytical calculations: in addition predictions from a numerical analysis are discussed.
\end{abstract}

\section{INTRODUCTION}

Ferrite kicker magnets are used extensively within the CERN accelerator complex to inject and extract beams from the various machines for transfer to experimental areas, other particle accelerators or to dump beams. They have been known to be a significant source of beam coupling impedance in the accelerator complex at CERN for some time \cite{benoit-thesis}. This is due to the proximity of very lossy materials to the beam. To counteract this a number of impedance reduction techniques have been applied to these magnets, both retroactively to existing pieces of equipment \cite{sps-mke-imp-red} and during the design stages \cite{mki-beam-screen}. 

Due to the difficulty of simulating ferrite kickers with the impedance reduction measures in place, it is often more convenient to measure their beam impedance using a coaxial wire setup as opposed to using simulations. Further work can subsequently be done to understand the measurements and give guidance to simulation studies.

\section{COAXIAL WIRE MEASUREMENT METHODS}


The coaxial wire method has been used for a number of decades as a bench-top method of measuring the beam coupling impedance of accelerator structures. It is based on the concept that the electromagnetic field profile around an ultrarelativistic charged particle is similar in nature to that of a short electrical pulse propagating along a coaxial line. 

For the measurements of the LHC-MKI (Large Hadron Collider Injection Kicker Magnets) the resonant coaxial wire method has been used. For the other measurements presented here the classical transmission method has been used. Both of these are described in more detail in \cite{sps-mke-imp-red, day-coaxial}.

\section{LHC-MKI INJECTION KICKER MAGNETS}

The LHC-MKIs are a set of transmission line ferrite kicker magnets. Unlike most existing kicker magnets, they were designed with impedance reduction mechanisms in place, in the form of a ceramic tube within which a series of longitudinal (from the beam point of view) conducting strips are inserted to provide a low resistivity path for the image currents. This was done to reduce both beam induced heating \cite{kicker-heating} and the likelihood of impedance based instabilities in the circulating beam. 

\subsection{Longitudinal Impedance}

The longitudinal impedance of the LHC-MKI was measured using the resonator method. For reference measurements, the beampipe of the counter-rotating beam in the kicker assembly was used. This is a copper tube of equal length to the main magnet aperture. Values without the beam screen are given by the Tsutsui formalism for two parallel plates \cite{tsutsui-long,tsutsui-dip,tsutsui-quad}.

\begin{figure}
\begin{center}
\includegraphics[width=0.8\linewidth]{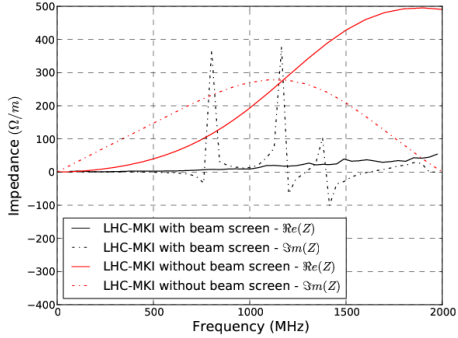}
\end{center}
\caption{The longitudinal impedance per unit length of the LHC-MKI with and without the beam screen.}
\label{fig:mki-long}
\end{figure}

\begin{figure}
\begin{center}
\includegraphics[width=0.8\linewidth]{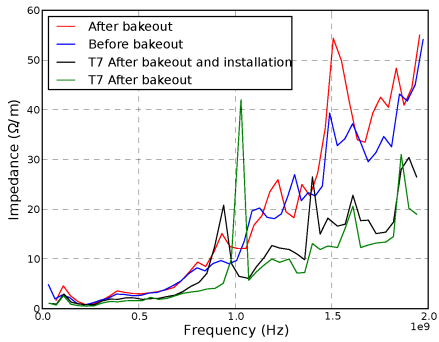}
\end{center}
\caption{The real longitudinal impedance per unit length of the LHC-MKI compared at different stages. T7 (Tank7) is separate magnet from the other measurements.}
\label{fig:mki-long-comp}
\end{figure}

As can be seen in Fig.~\ref{fig:mki-long}, the inclusion of the beam screen is highly effective in reducing the real impedance of the magnet. The imaginary impedance is also greatly reduced across the majority of the measured frequency spectrum, however two large peaks occur at $\sim$800 MHz and $\sim$1200 MHz. A series of measurements were also done on an LHC-MKI magnet at different stages of preparation to see how operation may condition the impedance of the device. Fig.~\ref{fig:mki-long-comp} demonstrates that although the beam screen remains effective in suppressing the real impedance, the preparations for insertion in the LHC and operation with the LHC can cause significant changes in the impedance of devices built to the same specification, particularly at frequencies greater than 1 GHz. Simulations to understand the source of these peaks are currently under investigation.


\subsection{Transverse Impedance}

\begin{figure}
\begin{center}
\includegraphics[width=0.8\linewidth]{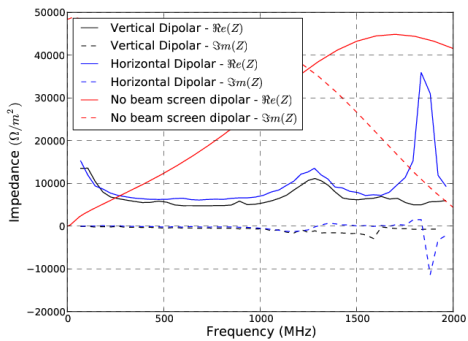}
\end{center}
\caption{The dipolar impedance per unit length of the LHC-MKI with and vertical dipolar without the beam screen.}
\label{fig:mki-dipolar}
\end{figure}
\begin{figure}
\begin{center}
\includegraphics[width=0.8\linewidth]{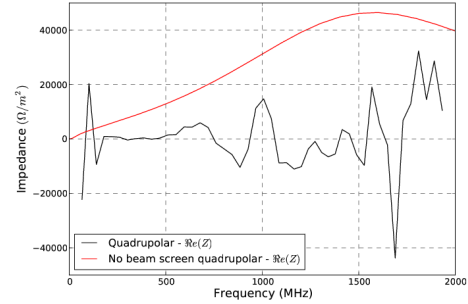}
\end{center}
\caption{The real horizontal quadrupolar impedance per unit length of the LHC-MKI with and without the beam screen.}
\label{fig:mki-quadrupolar}
\end{figure}

In common with the longitudinal impedance, the transverse (both dipolar and quadrupolar) impedances are greatly reduced when compared to that expected for unscreened ferrite. The large peak observed in the horizontal dipolar impedance at 1800 MHz (see Fig.~\ref{fig:mki-dipolar}) is thought to be a Higher Order Mode (HOM) present in the measuring setup and thus does not represent a real impedance. The quadrupolar impedance is drastically reduced at all frequencies by the beam screen. The behaviour of the transverse impedances with respect to the state of the beam screen is similarly under investigation using simulation models. 

\section{SPS-MKE EXTRACTION KICKER MAGNETS}

\begin{figure}
\begin{center}
\includegraphics[width=0.8\linewidth]{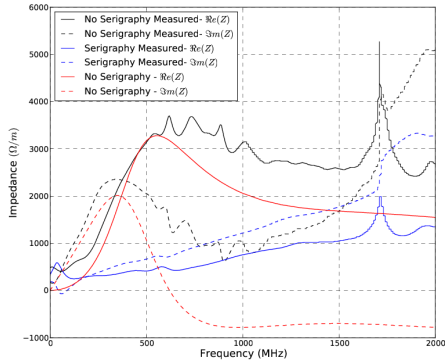}
\end{center}
\caption{The longitudinal impedance per unit length of the SPS-MKE with serigraphy (measured) and without serigraphy (measured and calculated using the Tsutsui model).}
\label{fig:mke-long}
\end{figure}

The SPS-MKEs (extraction kicker magnet) are transmission line ferrite kicker magnets to allow the extraction of beam from the SPS to the LHC. Two families of MKEs exists; the L-type and the S-type, differing in the dimensions of the kicker aperture. Originally designed without impedance reduction techniques in place, recent desires for increased beam performances were the cause for a concerted effort to devise effective retroactive impedance reductions techniques to be applied to these magnets \cite{sps-mke-imp-red}. Comparisons between the measured impedances of two magnets with and without the reduction, as well as an analytical calculation with the Tsutsui model can be seen in Fig.~\ref{fig:mke-long}. The impedance of the SPS-MKE was measured using the classical transmission method. For the reference measurement an analytical calculation was made.

The longitudinal impedance of the SPS-MKE is greatly suppressed by the inclusion of serigraphy on the surface of the magnet (Fig.~\ref{fig:mke-long}). Furthermore a comparison of measurements of the magnets without serigraphy to the Tsutsui model indicate a very good agreement up to 600 MHz, with divergence beyond this frequency due to differences between the internal structure of the magnet and the Tsutsui model. Again the large peak at $\sim$1700 MHz is postulated to be an artifact of the measurement.
\section{SPS-MKP INJECTION KICKER MAGNETS}

\begin{figure}
\begin{center}
\includegraphics[width=0.8\linewidth]{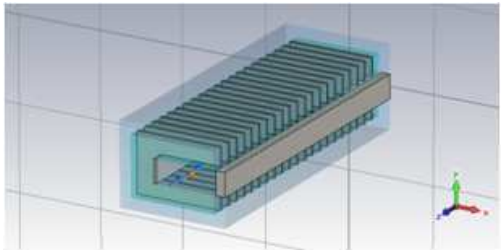}
\end{center}
\caption{A cut away of the SPS-MKP model. Note the thin layers of segmentation between ferrite blocks. The model is simulated with a mesh count of 1.3 million.}
\label{fig:mkp-geo}
\end{figure}

The SPS-MKP (injection kicker magnet) are a set of transmission line magnets used to inject beam into the SPS. One of their features is that they are composed of a many segments of ferrite separated by high voltage plates as can be seen in Fig.~\ref{fig:mkp-geo}. Measurements and simulations have been carried out to determine how the inclusion of the segmentation causes the measurements to differ from the analytic models.The impedance of the SPS-MKP was measured using the classical transmission method. Simulations were carried using CST Particle Studio \cite{cst-ref}.

\begin{figure}
\begin{center}
\includegraphics[width=0.8\linewidth]{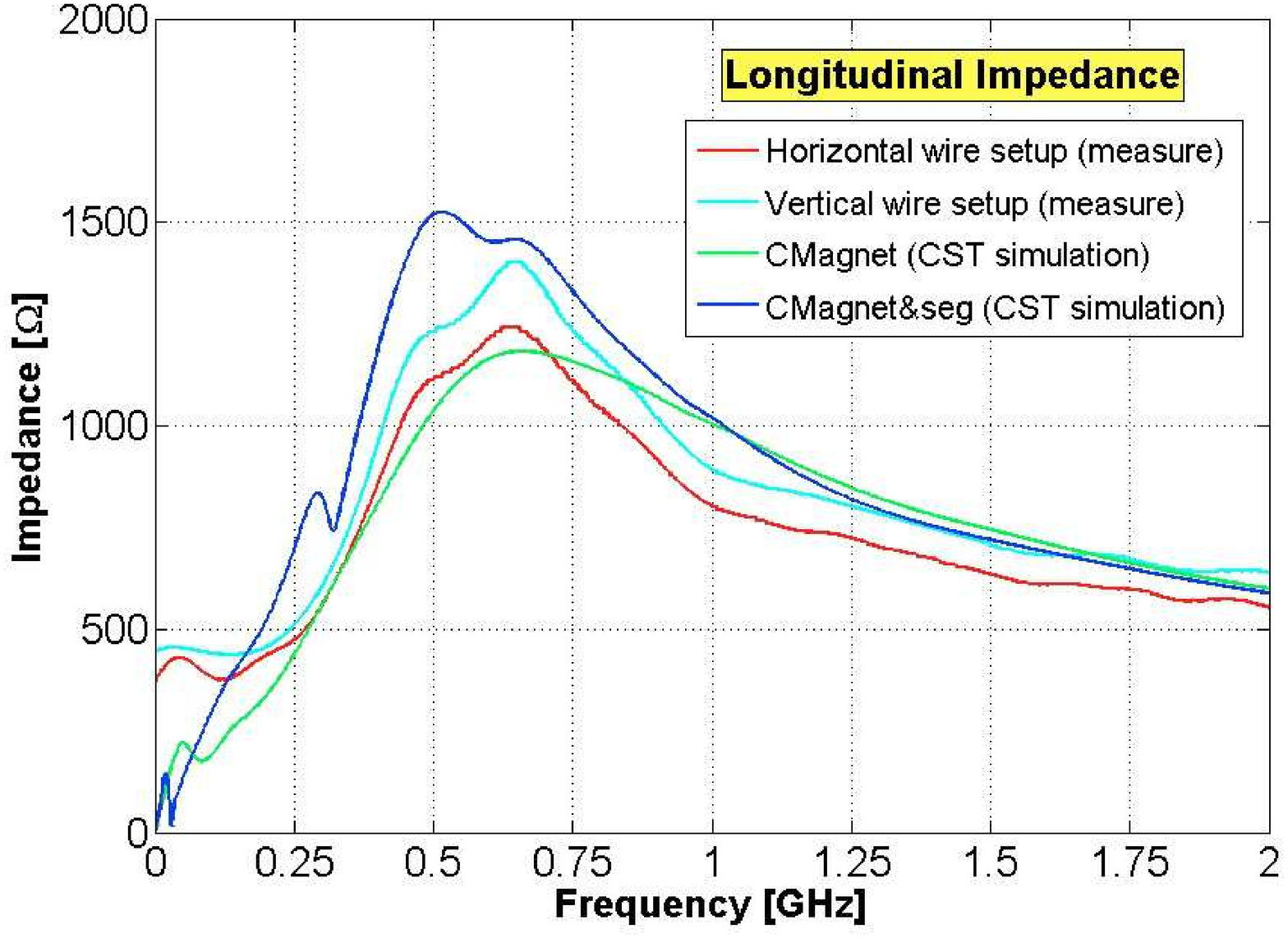}
\end{center}
\caption{The real longitudinal impedance of the SPS-MKP. Comparison between measurements and simulations with and without the segmentation. Seg indicates simulations with segmentation. Horizontal/vertical measure refer to measurements made during displaced with measurements in the horizontal and vertical plane respectively.}
\label{fig:mkp-long}
\end{figure}

As can be seen from the simulations (see Fig.~\ref{fig:mkp-long}), the addition of segmentation and the C-core shape cause an increase in the peak value of the real impedance at $\sim$600 MHz and also the generation of a lower frequency peak at 30-40 MHz. 

\section{CONCLUSION AND OUTLOOK}
We have demonstrated the effectiveness of the implemented impedance reduction techniques in lowering the beam coupling impedance of a number of ferrite kicker magnets at the CERN accelerator complex. In particular we have a significant reduction in the real impedance compared to their unshielded cases thereby reducing the problem of beam-induced heating. Furthermore we can see the theoretical models currently in use are not sufficient to fully explain the measured impedance profiles due to not truly representing the internal magnet structure. Further work is underway to develop models that more accurately represent the magnets internal structure.

\section{ACKNOWLEDGEMENTS}
Thanks to Yves Sillanoli and Salim Bouleghlimat for help in the mechanical setup of the measurement system.

\end{document}